\documentstyle[epsfig,prb,aps,multicol]{revtex}
\begin{document}
\draft
\title{Effects of the field modulation on the Hofstadter's spectrum}
\author{Gi-Yeong Oh$^{*}$}
\address{Department of Basic Science, Ansung National University,}
\address{Kyonggi-do 456-749, Korea}
\date{\today}
\maketitle

\begin{abstract}
We study the effect of spatially modulated magnetic fields on the
energy spectrum of a two-dimensional (2D) Bloch electron. Taking
into account four kinds of modulated fields and using the method
of direct diagonalization of the Hamiltonian matrix, we calculate
energy spectra with varying system parameters (i.e., the kind of
the modulation, the relative strength of the modulated field to
the uniform background field, and the period of the modulation)
to elucidate that the energy band structure sensitively depends
on such parameters: Inclusion of spatially modulated fields into
a uniform field leads occurrence of gap opening, gap closing, band
crossing, and band broadening, resulting distinctive energy band
structure from the Hofstadter's spectrum. We also discuss the
effect of the field modulation on the symmetries appeared in the
Hofstadter's spectrum in detail.
\end{abstract}

\begin{multicols}{2}
\section{Introduction}

The problem of a 2D Bloch electron under a uniform magnetic field
has been intensively studied for several decades,$^{1}$ and it is
well known that the energy spectrum is characterized by the
Hofstadter's butterfly showing a fractal nature.$^{2}$ Recently
the problem has attracted theoretically renewed interest in
connection with various phenomena such as the quantum Hall
effect,$^{3}$ the flux state model for high-$T_c$
superconductivity,$^{4}$ and the mean-field transition temperature
of superconducting networks or Josephson junction arrays.$^{5}$
Besides, recent advance in submicron technology that makes it
possible to fabricate any desired microstructures has stimulated
experimental study to find indication of the Hofstadter's spectrum
and its effect on the transport and optical properties.$^{6-9}$ In
parallel with this problem, the study on 2D electron systems under
nonuniform (either disordered$^{10}$ or periodic$^{11-26}$)
magnetic fields has been extensively performed, and lots of
interesting characteristics in the energy spectral and transport
properties have been elucidated.

Though the problem of a 2D Bloch electron under spatially
modulated magnetic fields has attracted less attention compared
with the problem of 2D electron gas under spatially modulated
magnetic fields,$^{11-22}$ it is still an important problem not
only in the view of theoretical interest but also in the view of
experimental interest, and there have been attempts to solve this
problem.$^{23-26}$ However, unfortunately, some of the relevant
works contain inconsistent results on the energy spectral
properties: In Ref.~24, Gumbs {\it et al}. studied the effect of a
one-dimensional sine-modulated (1DSM) field on a 2D Bloch electron
to argue that the symmetries appeared in the Hofstadter's spectrum
break down by the field modulation. Most surprisingly, they also
argued that the field modulation leads an additional crisscross
pattern like a spiderweb structure onto the Hofstadter's spectrum.
However, Oh {\it et al}.$^{25}$ studied the effect of a 1D
cosine-modulated (1DCM) field on a 2D Bloch electron to elucidate
that occurrence of gap closing and gap opening leads different
energy spectrum from the Hofstadter's spectrum and that the
symmetries appeared in the Hofstadter's spectrum except the dual
property still remain despite the field modulation. In the
meanwhile, Shi and Szeto$^{26}$ studied the energy spectrum of a
2D Bloch electron under a kind of 2D field modulation to argue
that there is no symmetry breaking in the energy spectrum and that
the fractal structure remains irrespective of the field
modulation.

In this paper we reexamine the problem of a 2D Bloch electron
under spatially modulated magnetic fields to settle the
inconsistency discussed above. In doing this, we take into account
four kinds of modulated fields (i.e., 1DSM, 1DCM, 2DSM, and 2DCM
fields) in order to obtain rather generic effects of the field
modulation on the Hofstadter's spectrum. By means of direct
diagonalization of the Hamiltonian matrix, we calculate the energy
eigenvalues and examine how the system parameters such as the kind
of the modulation, the relative strength of the modulated field to
the uniform field, and the period of the modulation influence on
the energy band structure and the symmetry of the Hofstadter's
spectrum. Introduction of the field modulation is shown to change
the $\vec{k}$ dependence of the energy spectrum drastically,
leading occurrence of gap opening, gap closing, band crossing, and
band broadening, which are the origin of distinctive energy band
structure from the Hofstadter's spectrum. Our results indicate
that there is no additional spiderweb structure in the energy
spectrum contrary to the result of Ref.~24 and that the field
modulation generically breaks the symmetries and the fractal
property of the Hofstadter's spectrum.

This paper is organized as follows: In Sec.~II we introduce four
kinds of magnetic fields and the tight-binding model. And, in
Sec.~III we present numerical results on the energy band structure
of a Bloch electron and the symmetries of the energy spectrum
under these fields. Section~IV is devoted to a brief summary.

\section{Modulated magnetic fields and the tight-binding model}

We consider an electron in a 2D square lattice under a spatially
modulated magnetic field
\begin{equation}
\vec{B}=\left[B_{0}+B_{1}(x,y)\right]\hat{z},
\end{equation}
where $B_{0}~(B_{1})$
denotes the uniform (modulated) part of an applied magnetic field.
Among possible kinds of modulated fields, we pay attention to two
kinds of modulated fields. The one is the SM field
\begin{equation}
  B_{1}(x,y)=B_{x}\sin\left(\frac{2\pi x}{T_{x}}\right)
            +B_{y}\sin\left(\frac{2\pi y}{T_{y}}\right) \label{eq:1}
\end{equation}
and the other is the CM field
\begin{equation}
  B_{1}(x,y)=B_{x}\cos\left(\frac{2\pi x}{T_{x}}\right)
            +B_{y}\cos\left(\frac{2\pi y}{T_{y}}\right).\label{eq:2}
\end{equation}
Here $B_{x(y)}$ is the strength of the modulated field and
$T_{x(y)}$ is the period of the modulation along the $x$ $(y)$
direction. Under the Landau gauge, the vector potential becomes
\begin{eqnarray}
A_{x}&=&\frac{B_{y}T_{y}}{2\pi}
              \cos{\left(\frac{2\pi y}{T_{y}}\right)}, \nonumber\\
A_{y}&=&B_{0}x-\frac{B_{x}T_{x}}{2\pi}
              \cos{\left(\frac{2\pi x}{T_{x}}\right)}  \label{eq:3}
\end{eqnarray}
for the SM field and
\begin{eqnarray}
A_{x}&=&-\frac{B_{y}T_{y}}{2\pi}
             \sin{\left(\frac{2\pi y}{T_{y}}\right)}, \nonumber\\
A_{y}&=&B_{0}x+\frac{B_{x}T_{x}}{2\pi}
             \sin{\left(\frac{2\pi x}{T_{x}}\right)}  \label{eq:4}
\end{eqnarray}
for the CM field, respectively.

The tight-binding Hamiltonian describing the motion of an electron
in a magnetic field is given by
\begin{equation}
 H=-\sum_{ij}t_{ij}e^{i\theta_{ij}}|i\rangle\langle j|, \label{eq:5}
\end{equation}
where $t_{ij}$ is the hopping integral between the
nearest-neighboring sites $i$ and $j$, and
$\theta_{ij}\equiv (2\pi e/hc)\int_{i}^{j}\vec{A}\cdot d\vec{l}$
is the magnetic phase factor. Under the vector potentials given by
Eqs.~(\ref{eq:3}) and (\ref{eq:4}), the magnetic phase factor becomes
\begin{equation}
 \theta_{mn;m^{'}n^{'}}=\left\{ \begin{array}{ccl}
         \pm\theta_{m} &,& (m^{'},n^{'})=(m,n\pm 1) \\
         \pm\theta_{n} &,& (m^{'},n^{'})=(m\pm 1,n) \\
         0 &,& \mbox{otherwise} \end{array}\right.     \label{eq:6}
\end{equation}
with
\begin{eqnarray}
 \theta_{m}&=& 2\pi m\phi_{0}-\beta_{x}\gamma_{x}\phi_{0}\cos
    \left(\frac{2\pi m}{\gamma_{x}}\right), \nonumber \\
 \theta_{n}&=&\beta_{y}\gamma_{y}\phi_{0}\cos
    \left(\frac{2\pi n}{\gamma_{y}}\right)             \label{eq:7}
\end{eqnarray}
for the SM field and
\begin{eqnarray}
 \theta_{m}&=& 2\pi m\phi_{0}+\beta_{x}\gamma_{x}\phi_{0}\sin
    \left(\frac{2\pi m}{\gamma_{x}}\right), \nonumber \\
 \theta_{n}&=&-\beta_{y}\gamma_{y}\phi_{0}\sin
    \left(\frac{2\pi n}{\gamma_{y}}\right)             \label{eq:8}
\end{eqnarray}
for the CM field, respectively. Here $\beta_{x(y)}=B_{x(y)}/B_{0}$,
$\gamma_{x(y)}=T_{x(y)}/a$, and $\phi_{0}=B_{0}a^{2}$, $a$ being
the lattice constant. The magnetic flux per unit cell, in units of
the flux quantum $hc/e$, is given by
$\phi=(1/2\pi)\sum\theta_{ij}=\oint\vec{A}\cdot d\vec{l}
=\int\vec{B}\cdot d\vec{S}$.

By means of Eqs.~(\ref{eq:5}) and (\ref{eq:6}), the tight-binding
equation can be written as
\begin{eqnarray}
 &&e^{i\theta_{n}}\psi_{m+1,n}+e^{-i\theta_{n}}\psi_{m-1,n}
 +\lambda\left( e^{i\theta_{m}}\psi_{m,n+1}\right. \nonumber \\
 &&\left.+e^{-i\theta_{m}}\psi_{m,n-1}\right)= E\psi_{mn} \label{eq:9}
\end{eqnarray}
where $\lambda$ $(\equiv t_{y}/t_{x})$ is the ratio of hopping
integrals between the $x$ and $y$ directions, and $E$ is the energy
in units of $t_{x}$. Here, the wave function is given by
$|\psi\rangle=\sum_{j}\psi_{j}|j\rangle$.

Denoting $R_{x(y)}$ as the periodicity of $\theta_{m(n)}$, the
Bloch theorem can be written as
\begin{equation}
\psi_{m+R_{x},n}=e^{ik_{x}R_{x}}\psi_{mn},~~
\psi_{m,n+R_{y}}=e^{ik_{y}R_{y}}\psi_{mn} ,           \label{eq:10}
\end{equation}
and the first magnetic Brillouin zone (FMBZ) is given by
$|k_{x(y)}|\leq\pi/R_{x(y)}$. We calculate the energy eigenvalues
for all the values of $\vec{k}$ in the FMBZ by directly
diagonalizing the Hamiltonian matrix obtained from
Eqs.~(\ref{eq:9}) and (\ref{eq:10}).

\section{Numerical Results and Discussion}

In what follows, we assume the modulated field has a
square-symmetry (i.e., $\beta_{x}=\beta_{y}=\beta$ and
$\gamma_{x}=\gamma_{y}=\gamma$) and consider only the case of the
isotropic hopping integral (i.e., $\lambda=1$) for the sake of
simplicity. And we pay attention to the energy dispersions for
$q=2,3$ (with setting $p=1$) and $\gamma=2,3,4$ because energy
dispersions for other values of $(q,\gamma)$ can be obtained in
a similar way. Here $p$ and $q$ denote the numbers (prime each
other) given by $\phi_{0}=p/q$.

\subsection{Energy spectrum in a uniform magnetic field}

In a uniform field ($\beta=0$), $\theta_{n}$ becomes zero. Thus,
by means of the translational invariance along the $y$ direction,
Eq.~(\ref{eq:9}) reduces to the Harper equation
\begin{equation}
 \psi_{m+1}+\psi_{m-1}+2\cos(\theta_{m}+k_{y})\psi_{m}
 =E\psi_{m},                                   \label{eq:11}
\end{equation}
and there are $q$ eigenvalues for a given value of $k_{y}$. When
$q$ is odd, the full energy spectrum consists of $q$ subbands. In
the meanwhile, when $q$ is even, the full energy spectrum consists
of $(q-1)$ subbands because two central subbands touch at zero
energy. Here the full energy spectrum means a set of energy
eigenvalues obtained by taking into account all the values of
$\vec{k}$ in the FMBZ. Energy dispersions for $q=2$ and $3$ are
given as follows:

(a) $q=2$: We have $R_{x}=2$ and the energy dispersion is given by
\begin{equation}
 E(\vec{k})=\pm 2\sqrt{\cos^{2}k_{x}+\cos^2 k_{y}}.  \label{eq:12}
\end{equation}
There are two subbands and they touch at zero energy as shown in
Fig.~1(a), resulting a single subband structure with $|E|\leq
2\sqrt{2}$.

(b) $q=3$: We have $R_{x}=3$ and the energy dispersion is given
by the equation
\begin{equation}
  E^{3}-6E-2(\cos 3k_{x}+\cos 3k_{y})=0.             \label{eq:13}
\end{equation}
There are three subbands as shown in Fig.~1(b). But, there is no
touching point in the dispersion and the full energy spectrum
exhibits a three subband structure with $-(1+\sqrt{3})\leq
E\leq-2$, $|E|\leq(\sqrt{3}-1)$, and $2\leq E\leq(1+\sqrt{3})$.

It is well known that the Hofstadter's spectrum has the following
symmetries; (i) the dual property between the $k_{x}$ and $k_{y}$
directions (let us denote it as $S_{D}$), (ii) the symmetry between
$-E$ and $E$ ($S_{E}$), (iii) the symmetry between $-k_{x}$ and
$k_{x}$ ($S_{X}$), and (iv) the symmetry between $-k_{y}$ and
$k_{y}$ ($S_{Y}$). It should be noted that $S_E$ holds only when
all the values of $\vec{k}$ in the FMBZ are taken into account;
for a given value of $\vec{k}$, one can easily check that $S_E$
does not hold for odd $q$.

\subsection{Energy spectrum in 1D modulated magnetic fields}

When the field modulation is along the $x$ direction, $\theta_{n}$
is still zero and the tight-binding equation is formally the same as
Eq.~(\ref{eq:11}). However, $\beta$ and $\gamma$ are introduced in
$\theta_{m}$, and $\phi$ becomes periodic along the $x$ direction
with the period $\gamma$:
\begin{equation}
 \phi/\phi_{0}=\left\{\begin{array}{l}
     1+(\beta\gamma/2\pi)\left[\cos\left(2\pi(m+1)/
     \gamma\right)\right. \\
     ~~~~~~ -\left.\cos\left(2\pi m/\gamma\right)\right]
     ~~~~~~~~{\rm for}~~\mbox{1DSM} \\
     1-(\beta\gamma/2\pi)\left[\sin\left(2\pi(m+1)/
     \gamma\right)\right. \\
     ~~~~~~ -\left.\sin\left(2\pi m/\gamma\right)\right]
     ~~~~~~~~{\rm for}~~\mbox{1DCM}.
  \end{array}\right.                                \label{eq:15}
\end{equation}
The lattice is called the stripped flux lattice$^{23}$ and the
energy spectrum can be obtained by diagonalizing the $(R_{x}\times
R_{x})$ Hamiltonian matrix.

\subsubsection{1DSM Field}

When $(q,\gamma)=(2,2)$, we have $R_{x}=2$ and the energy dispersion
is given by
\begin{equation}
 E(\vec{k})=2\sin\beta\sin k_{y}\pm 2\sqrt{\cos^{2}k_{x}
 +\cos^{2}\beta\cos^{2}k_{y}}.                        \label{eq:16}
\end{equation}
The two subbands touch at $(k_{x},k_{y})=(\pi/2,\pi/2)$ and
$(\pi/2,3\pi/2)$, and the energy eigenvalues at the touching points
are given by $E=\pm2\sin\beta$. Since the touching points exist
irrespective of $\beta$, the full energy spectrum exhibits a single
subband structure as in the case of $\beta=0$.

In order to demonstrate the $\gamma$ dependence of the energy
spectrum with $q=2$, we calculate energy spectra for various
values of $\gamma$ and plot some of them in Fig.~2. Even though
there are six and four subbands for $(q,\gamma)=(2,3)$ and
$(2,4)$, there occurs direct touching between the nearest
neighboring subbands, which leads a single subband structure of
the full energy spectrum. In the calculations, we find that the
energy band structure is independent of the values of $\beta$ and
$\gamma$ while the total bandwidth slightly changes with varying
$\beta$.

Figure 3 shows the $k_y$ dependence of the energy spectrum for
$(q,\gamma)=(3,2)$. In this case, we have $R_{x}=6$ and there are
six subbands. For small values of $\beta$, the upper (lower) two
subbands touch at several points of $k_{y}$. Besides, there exists
indirect overlapping (i.e., crossing of subbands at different
values of $k_{y}$) between the central subbands. Thus, the full
energy spectrum still exhibits a three subband structure as can be
seen in Fig.~3(a). However, as $\beta$ increases, there occurs gap
opening between the subbands and the full energy spectrum exhibits
a six subband structure as in Fig.~3(b). Further increase of
$\beta$ makes the second and the fourth gaps closed, leading a
four subband structure as in Fig.~3(c). Figure 4 shows the $k_{y}$
dependence of the energy spectrum for $(q,\gamma)=(3,3)$. In this
case, we have $R_{x}=3$ and there are three subbands. For small
values of $\beta$, the full energy spectrum exhibits a three
subband structure as in Fig.~4(a). However, indirect overlapping
between subbands occurs as $\beta$ increases, and the full energy
spectrum exhibits a single subband structure as in Fig.~4(b).
Occurrence of gap closing due to indirect overlapping is also
found in the case of $(q,\gamma)=(3,4)$.

As for the symmetry of the energy spectrum, Figs.~2 and 3 show
that $S_E$ remains under the 1DSM field, which is contrary to the
arguments of Ref.~24. The reason of this discrepancy lies in the
range of $\vec{k}$ taken into account in discussing the symmetry
of the energy spectrum. In Ref.~24 only a particular value of
$\vec{k}$ was taken into account, while all the values of
$\vec{k}$ in the FMBZ are taken into account in this paper. Here
we would like to stress that all the values of $\vec{k}$ in the
FMBZ should be considered in order to discuss the symmetry of the
energy spectrum. This is because $S_E$ of the Hofstadter's
spectrum holds only when the all the values of $\vec{k}$ in the
FMBZ is considered. We also find that $S_X$ still remains under
the 1DSM field while whether $S_Y$ remains or not depends
crucially on $q$ and $\gamma$. Note that $S_D$ breaks down by
introducing the 1D field modulation.

\subsubsection{1DCM Field}

In order to test how the energy spectrum is influenced on the kind
of the field modulation, we also calculate the energy spectrum of
a 2D Bloch electron under the 1DCM field. Figure 5 shows the $k_y$
dependence of the energy spectrum for $(q,\gamma)=(2,3)$. In this
case, we have $R_{x}=6$ and there are six subbands. For small
values of $\beta$, the two central subbands directly touch at two
points of $k_{y}$ and the upper (lower) two subbands indirectly
overlap with each other, resulting a single subband structure as
in Fig.~5(a). However, as $\beta$ increases, indirect overlapping
between the upper (lower) two subbands disappears and there occurs
gap opening between them. Thus the full energy spectrum exhibits a
three subband structure as in Fig.~5(b). Further increase of
$\beta$ makes the remaining indirect overlapping disappear and the
full energy spectrum exhibits a five subband structure as in
Fig.~5(c). The $\beta$ dependence of the energy spectrum for
$(q,\gamma)=(2,4)$ is quite different from $(q,\gamma)=(2,3)$. In
this case, we have $R_{x}=4$ and there are four subbands. Each
subband directly touches with its neighboring subbands and the
full energy spectrum exhibits a single subband structure
regardless of $\beta$; the only effect of $\beta$ is to change the
energy bandwidth.

We also perform similar calculations for the energy spectrum with
$q=3$. When $(q,\gamma)=(3,3)$, we have $R_{x}=3$ and there are
three subbands. For small values of $\beta$, the three subbands
separate each other, exhibiting a three subband structure as in
the case of $\beta=0$. However, as $\beta$ increases, each subband
overlaps indirectly with its neighboring subbands and the gaps
between the subbands become closed, resulting a single subband
structure. We also find similar phenomena for $(q,\gamma)=(3,4)$.

In Table~1, we summarize the number of subbands of the full energy
spectra for the parameters $(q,\gamma,\beta)$ we took into
account. As for the symmetry of the energy spectrum, we find that
$S_E$, $S_X$, $S_Y$ remain, while $S_D$ breaks down under the 1DCM
field. Note that $S_Y$ depends on the kind of the field
modulation; it remains (breaks) in the 1DCM (1DSM) field.

Before ending this subsection, we would like to mention two
comments. First, we assume $\gamma$ to be an integer for the sake
of simplicity even though it can be arbitrary real value. Besides,
we consider only the values of $\gamma\geq 3$ ($2$) in the case of
the CM (SM) field because $\phi$ becomes nonuniform only under
these conditions. However, the authors of Ref.~24 violated this
condition; they chose $\gamma=1$ in their calculations. But, when
$\gamma=1$, $\phi$ becomes uniform as can be easily checked by
Eq.~(\ref{eq:15}) and the energy spectrum should be identical with
the Hofstadter's spectrum if all the values of $\vec{k}$ in the
FMBZ are taken into account, which implies that the results of the
cited paper, i.e., the appearance of additional crisscross pattern
like a spiderweb structure and the symmetry breaking of the energy
spectrum, are erroneous ones coming from mistakes in choosing the
values of $\gamma$ and in choosing the values of $\vec{k}$. Note
that the method of diagonalization of the Hamiltonian matrix used
in the present paper or in the cited paper and the method of the
transfer matrix used in Ref.~25 are equivalent. Actually,
Fig.~5(a) in the present paper is exactly the same as Fig.~2(b) in
Ref.~25. Second, the tight-binding model we are considering is
basically a one-band model. Thus, we focus our attention on energy
spectra only for the values of $\beta$ that are not large (i.e.,
$\beta\leq 1$) since there might be interband mixing between
different Landau levels for large values of $\beta$.

\subsection{Energy spectrum in 2D modulated magnetic fields}

When the field modulation is along both lateral directions,
$\theta_{n}$ becomes nonzero and $\phi$ becomes periodic in both
lateral directions with the period $\gamma$. The lattice is called
the checkerboard flux lattice$^{23}$ and the energy spectrum can
be obtained by diagonalizing the $(R_{x}R_{y}\times R_{x}R_{y})$
Hamiltonian matrix.

\subsubsection{2DSM field}

When $(q,\gamma)=(2,2)$, we have $(R_{x},R_{y})=(2,2)$ and the
Hamiltonian matrix becomes
\begin{equation}
 \left(\begin{array}{cccc}
   0   &   a   &   b   &   0 \\
 a^{*} &   0   &   0   &   c \\
 b^{*} &   0   &   0   &   d \\
   0   & c^{*} & d^{*} &   0
 \end{array}\right)                                  \label{eq:17}
\end{equation}
where $a=-e^{i\beta}-e^{-i(\beta+2k_{y})}$, $b=e^{-i\beta}
+e^{i(\beta-2k_{x})}$, $c=e^{i\beta}+e^{-i(\beta+2k_{x})}$, and
$d=e^{-i\beta}+e^{i(\beta-2k_{y})}$, respectively. By
diagonalizing Eq.(\ref{eq:17}) we obtain the energy dispersion as
\begin{eqnarray}
 E(\vec{k})=\pm 2\left| \cos\beta\sqrt{\cos^{2}k_{x}+\cos^{2}k_{y}}
 \nonumber \right. \\
 \left.\pm\sin\beta\sqrt{\sin^{2}k_{x}+\sin^{2}k_{y}}
 \right|,                                            \label{eq:18}
\end{eqnarray}
which is plotted in Fig.~6. Numbering the four subbands in order
of lowering energy, one can see that the first (second) and the
third (fourth) subbands directly touch at the center of FMBZ
regardless of $\beta$. There also exists band crossing between the
second and the third subbands, which is absent in the case of the
1D field modulation. Equation~(18) indicates that the full energy
spectrum exhibits a single subband structure regardless of $\beta$
and the total bandwidth changes with varying $\beta$.

When $(q,\gamma)=(2,3)$, we have $(R_{x}, R_{y})=(6,3)$ and there
are eighteen subbands. However, due to occurrence of direct
touching and indirect overlapping between neighboring subbands,
the full energy spectrum exhibits a single subband structure
regardless of $\beta$. When $(q,\gamma)=(2,4)$, the situation
becomes different. In this case, we have $(R_{x},R_{y})=(4,4)$
and there are sixteen subbands. For small values of $\beta$, due
to direct touching between neighboring subbands, the full energy
spectrum exhibits a single subband structure. However, gap opening
between subbands occurs with increasing $\beta$, resulting a five
subband structure of the full energy spectrum.

When $(q,\gamma)=(3,2)$, we have $(R_{x}, R_{y})=(6,2)$ and there
are twelve subbands. In this case, even for small values of
$\beta$, the full energy spectrum is quite different from the case
of $\beta=0$; the full energy spectrum exhibits a eight subband
structure as can be seen in Fig.~7(a). And, as $\beta$ increases,
two more gaps are open to yield a ten subband structure as in
Fig.~7(b). Further increase of $\beta$ makes the full energy
spectrum exhibit a twelve subband structure as in Fig.~7(c). When
$(q,\gamma)=(3,3)$, the effect of $\beta$ is quite different from
the case of $(q,\gamma)=(3,2)$. In this case, we have
$(R_{x},R_{y})=(3,3)$ and there are nine subbands. For small
values of $\beta$, the full energy spectrum consists of three
subbands as in the case of $\beta=0$. However, there occurs gap
closing with increasing $\beta$ and the energy spectrum exhibits a
single subband structure. We also find similar gap closing
behavior for $(q,\gamma)=(3,4)$.

\subsubsection{2DCM Field}

In order to know how the energy spectrum is influenced by the kind
of the 2D field modulation, we also calculate the energy spectra
of a Bloch electron under the 2DCM field with various values of
$(q,\gamma,\beta)$. And, we find that occurrence of gap closing,
gap opening, band crossing, and band broadening is a generic
effect of the field modulation.

When $(q,\gamma)=(2,3)$, there are eighteen subbands. For small
values of $\beta$, due to direct touching and indirect overlapping
between the inner sixteen subbands, the full energy spectrum
exhibits a three subband structure. However, there occurs gap
opening between the inner sixteen subbands with increasing
$\beta$, resulting a five subband structure. We also find that the
widths of gaps broaden with the increase of $\beta$. For
$(q,\gamma)=(2,4)$, there are sixteen subbands and they compose a
single subband structure for small values of $\beta$. However,
with increasing $\beta$, there occurs gap opening and the full
energy spectrum exhibits a five subband structure.

When $(q,\gamma)=(3,3)$, there are nine subbands and they compose
a three subband structure for small values of $\beta$. However,
gap closing occurs with increasing $\beta$, and the three subband
structure becomes a single subband structure. Further increase of
$\beta$ makes the full energy spectrum consist of seven subbands.
For $(q,\gamma)=(3,4)$, we show the $\beta$ dependence of the
energy spectrum in Fig.~8, where concurrent occurrence of gap
opening and gap closing can be seen. In this case, there are
twenty four subbands and they compose a three subband structure
for small values of $\beta$. However, as $\beta$ increases, there
occurs band splitting and the full energy spectrum exhibits a
seven subband structure. Further increase of $\beta$ makes the
full energy spectrum consist of three subbands. We summarize in
Table~2 the number of subbands of the full energy spectra for the
parameters $(q,\gamma,\beta)$ taken into account.

The symmetry of the energy spectrum under the 2D modulated fields
is rather complicated than the cases of the 1D modulated fields.
$S_E$ remains for both the 2DSM and 2DCM fields as can be seen in
Figs.~6--8. However, $S_X$ and $S_Y$ depend sensitively on the
system parameters. For the parameters considered, we find that
$S_X$ and $S_Y$ remain for the cases of $(q,\gamma)=(2,2), (2,4),
(3,2)$ in the 2DSM fields and $(q,\gamma)=(2,4), (3,3)$ in the
2DCM fields, while they break for the cases of $(q,\gamma)=(2,3),
(3,4)$ in the 2DSM fields and $(q,\gamma)=(2,3), (3,4)$ in the
2DCM fields. In the meanwhile, the energy spectrum in the 2DSM
field with $(q,\gamma)=(3,3)$ has a flipped symmetry with respect
to $k_{x(y)}=0$. Here the flipped symmetry means that the energy
spectrum in the range of $0\leq k_{x(y)}\leq\pi/R_{x(y)}$ is the
same as that in the range of $-\pi/R_{x(y)}\leq k_{x(y)}\leq 0$
when $E$ is replaced by $-E$. $S_D$ also depends sensitively on
the system parameters; it holds (breaks) for the values of
$(q,\gamma)$ that make $S_X$ and $S_Y$ remain (break).

Before ending up this subsection, we would like to mention a few
comments. The first is that the above results on $S_{E (X,Y,D)}$
are not from a theoretical analysis but from a numerical study.
Thus, further study such as the group theoretical
analysis$^{23,26,27}$ is required in order to deepen the
understanding of the symmetry breaking. The second is that the
$k_{x}$ and $k_{y}$ directions taken into account in discussing
$S_{X (Y,D)}$ are not the high symmetry directions of the
Hamiltonian under the 2D modulated fields. The reason for taking
into account these directions, nevertheless, lies in testing how
$S_{X (Y,D)}$ appeared in the Hofstadter's spectrum is influenced
by the field modulation. Since symmetry breaking is generally
expected if the symmetry axis is not properly chosen, our result
that the breaking of $S_{X (Y,D)}$ depends on the system
parameters is quite natural. Finally, it may be worthwhile to note
the result of Ref.~26, where Shi and Szeto considered a Bloch
electron in the magnetic field
\begin{equation}
 \vec{B}=[B_{0}+(-1)^{m-n}B_{1}]\hat{z}
\end{equation}
and found that there is no symmetry breaking in the energy
spectrum and that the fractal structure remains even though the
energy spectrum becomes different from the Hofstadter's spectrum,
which seems to be contrary to our results. However, since the
directions used in Ref.~26 are the highly symmetric $(k_{x}\pm
k_{y})$ directions while the directions used in this paper are the
$k_{x}$ and $k_{y}$ directions, it may not be easy to compare both
results directly. The only thing we can say is that the field
modulation given by in this paper [i.e., Eqs.~(1)-(3)] is more
generalized than that of Ref.~26 [i.e., Eq.~(19)]; specifying the
periodic magnetic field by a magnetic unit cell, the lattice with
Eq.~(19) becomes the simple checkerboard lattice [see, for
example, Fig.~2(a) of Ref.~23] while the lattice with Eqs.~(1)-(3)
becomes a generalized checkerboard lattice [see Fig.~2(b) of
Ref.~23]. Thus we expect that the latter may present more generic
effect of the field modulation than the former. Further
theoretical study on the comparison between two cases is also
required.

\section{Summary}

In summary, we studied the effects of spatially modulated magnetic
fields on the Hofstadter's spectrum. In order to obtain rather
general properties of the effect of the field modulation, we took
into account four kinds of modulated fields and calculated energy
eigenvalues with varying $q$, $\beta$, and $\gamma$. Occurrence of
gap opening, gap closing, band crossing, and band broadening was
found to be characteristic features of the field modulation, which
leads distinctive energy band structure from the Hofstadter's
spectrum. It should be noted that these characteristics are very
important since the change of the energy dispersion may have
crucial influence on the transport properties; it may allow an
electron to change its orbit, in view of the semiclassical
language. There is also a possibility of quenching the integer
quantum Hall behavior and exhibiting normal behavior when the
subgaps are smeared out. Sensitive dependence of the energy
spectrum on the system parameters implies that it may be difficult
to detect direct indication of the Hofstadter's spectrum in
experiment since even a tiny change of the magnetic field leads
much complicated energy band structure. We also studied the effect
of the field modulation on the symmetries appeared in the
Hofstadter's spectrum and illustrated that the symmetries
sensitively depend on the system parameters. In this work, we have
paid attention to the energy spectral properties only with a
rational flux and the isotropic hopping integral. Since an
introduction of hopping anisotropy is known to lead interesting
phenomena like band broadening,$^{28}$ it may be interesting to
study the effect of hopping anisotropy on the energy spectrum of a
2D Bloch electron under the modulated magnetic fields.


\vskip 0.4in
\centerline{\large\bf FIGURE CAPTIONS}
\vskip 0.2in

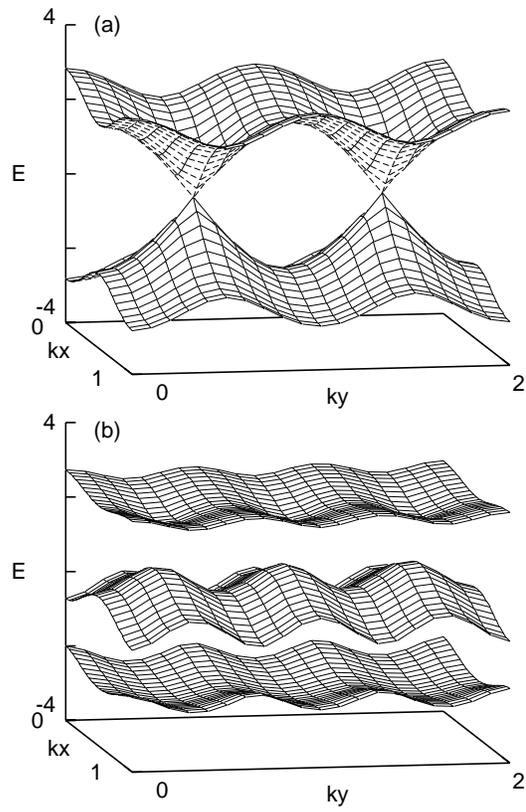
\noindent {\bf Fig.~1}
Energy dispersion in the uniform  magnetic field with (a) $q=2$
and (b) $q=3$. The horizontal plane is drawn in units of $\pi$.
\vskip 0.1in

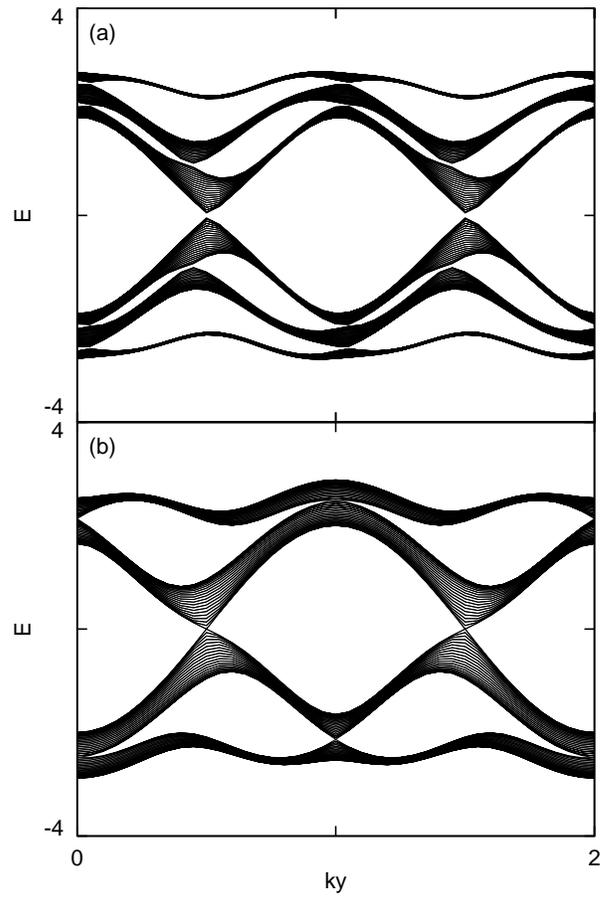
\noindent {\bf Fig.~2}
Plot of $E$ versus $k_{y}$ in the 1DSM field with $\beta=0.3$, where
(a) $(q,\gamma)=(2,3)$ and (b) $(q,\gamma)=(2,4)$. The horizontal
axis is drawn in units of $\pi$.
\vskip 0.1in

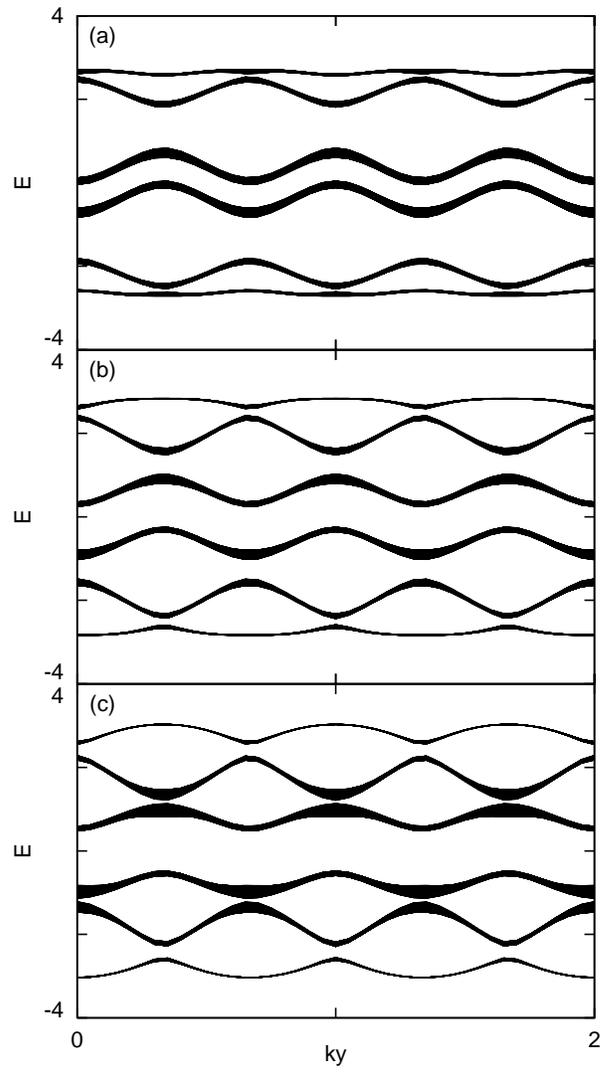
\noindent {\bf Fig.~3}
Plot of $E$ versus $k_{y}$ in the 1DSM field with $(q,\gamma)=(3,2)$,
where (a) $\beta=0.3$, (b) $\beta=0.6$, and (c) $\beta=0.9$.
\vskip 0.1in

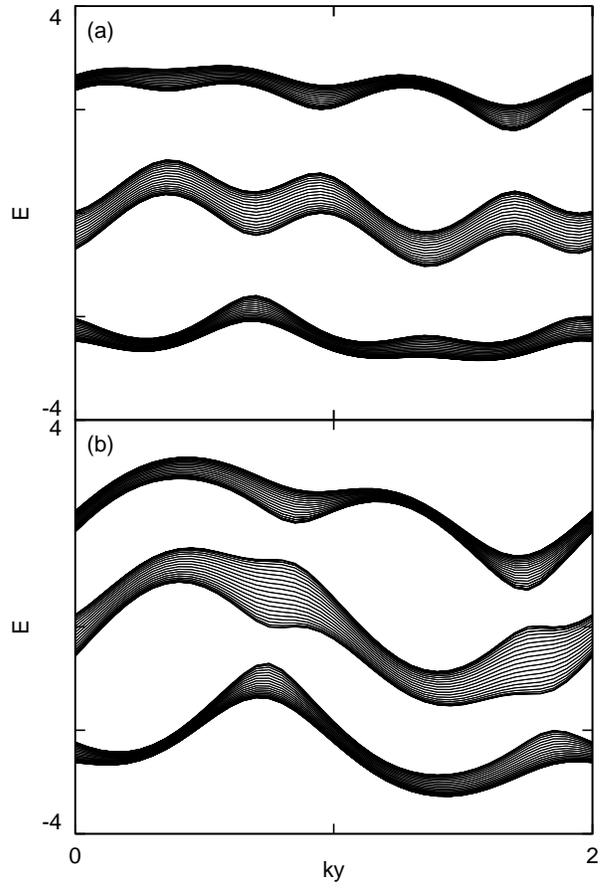
\noindent {\bf Fig.~4}
Plot of $E$ versus $k_{y}$ in the 1DSM field with $(q,\gamma)=(3,3)$,
where (a) $\beta=0.3$ and (b) $\beta=0.9$.
\vskip 0.1in

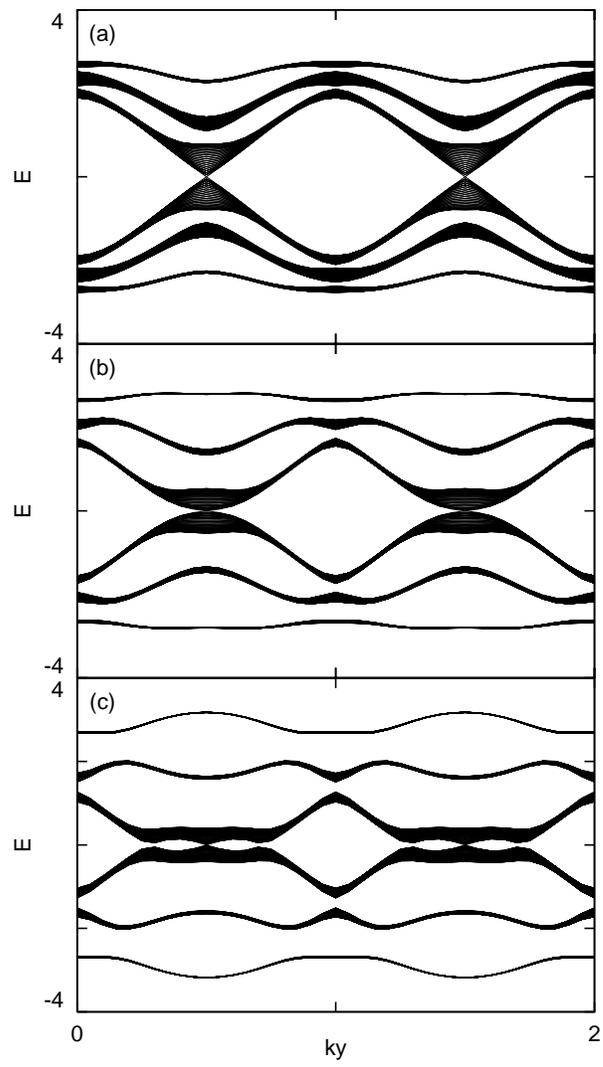
\noindent {\bf Fig.~5}
Plot of $E$ versus $k_{y}$ in the 1DCM field with $(q,\gamma)=(2,3)$,
where (a) $\beta=0.3$, (b) $\beta=0.6$, and (c) $\beta=0.9$.
\vskip 0.1in

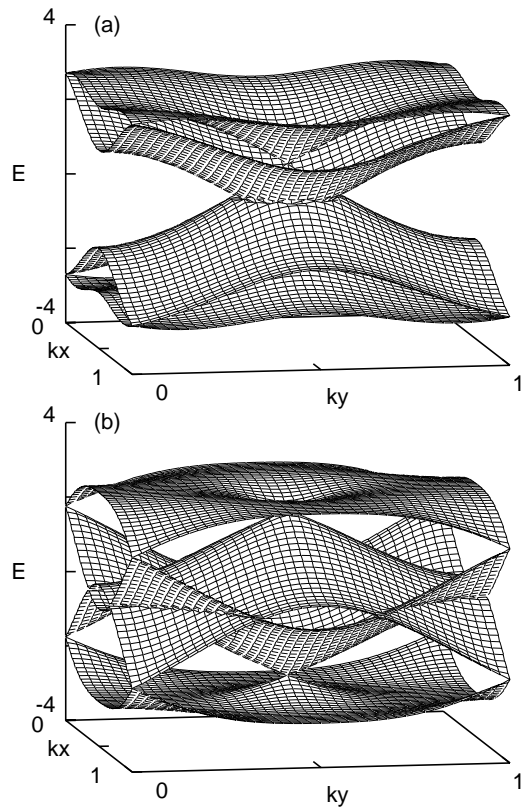
\noindent {\bf Fig.~6}
Energy dispersion in the 2DSM field with $(q,\gamma)=(2,2)$, where
(a) $\beta=0.3$ and (b) $\beta=0.9$.
\vskip 0.1in

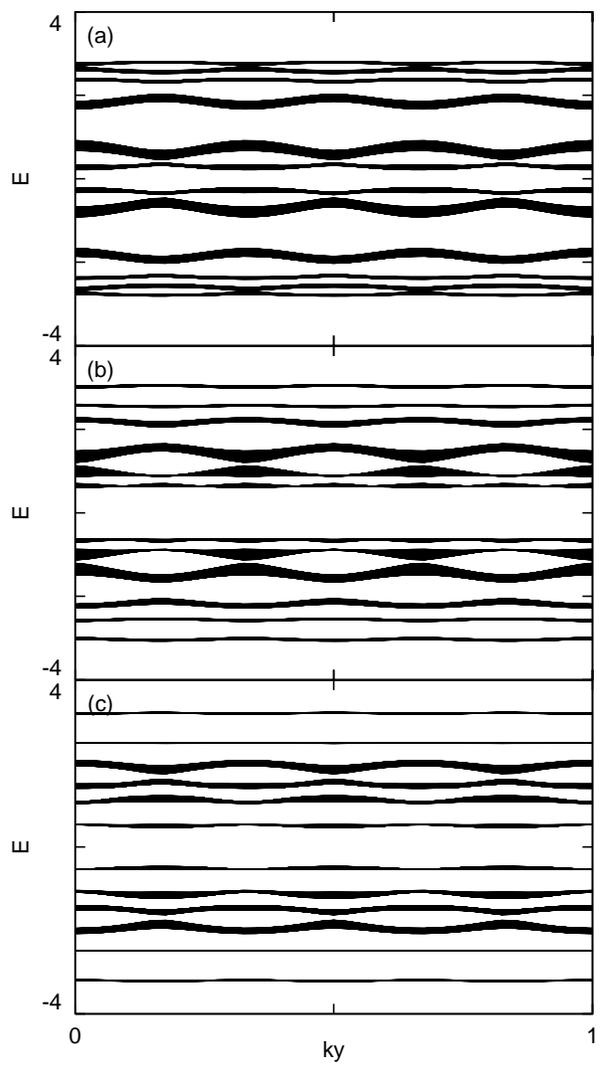
\noindent {\bf Fig.~7}
Plot of $E$ versus $k_{y}$ in the 2DSM field with $(q,\gamma)=(3,2)$,
where (a) $\beta=0.3$, (b) $\beta=0.6$, and (c) $\beta=0.9$.
\vskip 0.1in

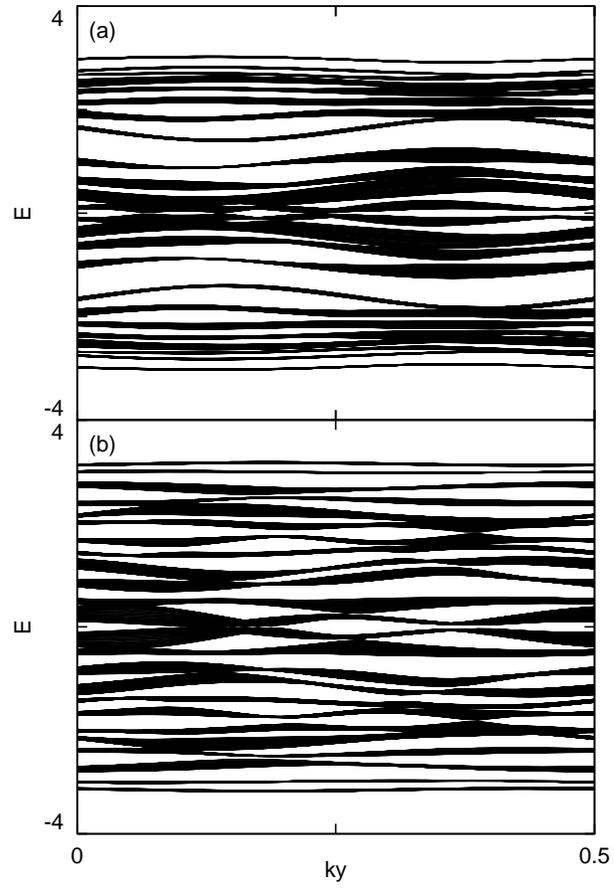
\noindent {\bf Fig.~8}
Plot of $E$ versus $k_{y}$ in the 2DCM field with $(q,\gamma)=(3,4)$,
where (a) $\beta=0.3$ and (b) $\beta=0.6$.

\vskip 0.4in
\centerline{\large\bf TABLE CAPTIONS}
\vskip 0.2in

\noindent {\bf Table~1}
Number of subbands of the full energy spectrum under the 1D modulated
fields.
\vskip 0.1in

\noindent {\bf Table~2}
Number of subbands of the full energy spectrum under the 2D modulated
fields.

\end{multicols}
\end{document}